\documentclass[aps,pre,preprint,showpacs,showkeys,superscriptaddress,nofootinbib,longbibliography,floatfix]{revtex4-1}
\usepackage[utf8]{inputenc}
\usepackage[english]{babel}
\usepackage{amsmath,amssymb,graphicx,bbm}
\usepackage{xcolor}
\usepackage{mathtools}
\usepackage[section]{placeins}

\newcommand{\mathe}{\mathrm{e}}
\begin{document}

\title{ Many-body methods in agent-based epidemic models}
 \author{Gilberto M. Nakamura}
 \email{gmnakamura@usp.br}
 \affiliation{Universidade de S\~{a}o Paulo, Ribeir\~{a}o Preto 14040-901, Brazil}
 \author{Alexandre S. Martinez}
 \email{asmartinez@ffclrp.usp.br}
\thanks{ We have applied techniques from fundamental physics to epidemic
  models, inspired by the teaching of Roger Maynard, to whom we dedicate
  our findings.  }
 \affiliation{Universidade de S\~{a}o Paulo, Ribeir\~{a}o Preto 14040-901, Brazil}
 \altaffiliation{Instituto Nacional de Ci\^{e}ncia e Tecnologia - Sistemas Complexos (INCT-SC)}
 \begin{abstract}
The susceptible-infected-susceptible (SIS) agent-based model is usually
employed in the investigation of epidemics. The model describes a Markov
process for a single communicable disease among susceptible (S) and
infected (I) agents.  However, the disease spreading forecasting is
often restricted to numerical simulations, while analytic formulations
lack both general results and perturbative approaches since they are
subjected to asymmetric time generators. 
Here, we discuss perturbation theory, approximations and application of
many-body techniques in epidemic models in the framework for
squared norm of probability vector $\vert P(t)\vert ^2$, in which 
asymmetric time generators are replaced by their symmetric counterparts.
 \end{abstract}

\maketitle

Proper planning and management lie in the foundation of efficient health
and sanitary policies~\cite{whoNEJM2016}. The decision-making process
usually relies on predictions from epidemic models and raw data to rule
the best available policy to mitigate the disease spreading. Resource
planning becomes even more relevant during the emergence of new
communicable diseases, as improper actions may extend the duration or
adversely impact health workers \cite{heesterbeekScience2015}.  Despite
the success attained by traditional epidemic models for large scale
epidemics in the past, they have been unable to produce reliable
predictions for small scale disease spreading in heterogeneous
populations~\cite{heesterbeekScience2015,bansalJRSoc2007,keelingPNAS2002}. This
issue is further enhanced due to the stochastic nature of pathogen
transmission mechanisms and patient care.  As such, a considerable
amount of epidemic models have been proposed to mimic the correct
behavior for spreading of communicable infectious diseases.

The simplest susceptible-infected-susceptible (SIS) model considers the
time evolution of a single communicable disease among $N$ susceptible
(S) and infected (I) agents~\cite{satorrasRevModPhys2015}.  The infected
agents may either transmit the disease to susceptible agents with
constant probability $\alpha$, turning them into infected agents
$S\rightarrow I$, or undergo the recovery process $I\rightarrow S $ with
probability rate $\beta$, during a fixed time interval $\delta
t$. Furthermore, two approaches are available to describe the disease
spreading in the SIS model for a fixed population of size $N$:
compartmental and stochastic.  In  the compartmental approach relevant
quantities are well-described by averages 
\cite{keelingJRSoc2005}, from which one derives non-linear differential equations. 
For instance, the number of infected agents $n(t)$, for
fixed $N$, in the compartmental SIS model is  
\begin{equation}
  \frac{d n}{dt} = \tilde{\alpha} n\left(1-\frac{n}{N}\right)- \tilde{\beta} n.
\label{eqcompartimental1}
\end{equation}
This is the expected behavior for large homogeneous populations, where
fluctuations are negligible.  Introduction of effective transmission
($\tilde{\alpha}$) and recovery ($\tilde{\beta}$) probabilities rates
contemplates effects due to heterogeneous population. This parameter
estimations employ networks, with size $N$, as substrate to display the
varying degree of non-uniformity within a population.  In this scheme,
each vertex in the network contains a single agent, while the links
among agents are given by the corresponding adjacency matrix $A$.  
Thus, non-trivial topological aspects of $A$ are incorporated in the
effective transmission and cure
rates~\cite{bansalJRSoc2007,keelingPNAS2002,satorrasRevModPhys2015,keelingJRSoc2005,satorrasPhysRevLett2001}.

The stochastic approach also requires networks to describe the
population heterogeneity.  However, contrary to the compartmental
approach, the assumption about negligible fluctuations is removed,
meaning averages alone are not sufficient to properly describe the
disease spreading.   In fact, fluctuations are intrinsic components in
stochastic formulations and their relevance increases with vanishing
$N$~\cite{vanKampen1981}, a much more realistic scenario in modeling
emergence of small scale epidemics of communicable infectious
diseases~\cite{heesterbeekScience2015}. In this approach, transition
probabilities among configurations of $N$ agents are expressed by the
transition matrix $\hat{T}$, and take place during the time interval
$\delta t$~\cite{reichl1998,alcarazAnnPhys1994,alcarazPhysRevE2008}. The
disease transmission and agent recovery are modeled as probability
vector $|P(t)\rangle$ of a Markov process with time interval $\delta t$
taken to be small enough so that only one recovery or one transmission
event takes place in the entire population.  This is compatible with the
Poissonian assumption~\cite{satorrasRevModPhys2015}.

As a closing remark, one notices the transition matrix is analogous to 
the time evolution operator in quantum theories. As a result, the
eigenvalues and eigenvectors of $\hat{T}$ express the time evolution of
agent-configurations.  Despite the striking similarity between both
operators, $\hat{T}$ is often asymmetric, restraining its use to small
values of $N$ for epidemic models in numerical simulations or
introducing severe obstacles for analytic solutions.  These issues
hinder the systematic development of perturbation theories for
agent-based epidemic models, often requiring fresh simulations to
forecast the impact of small variations of the parameters of the model,
contrary to the rationale behind perturbation theory.

Here, we first briefly review results  \cite{nakamuraArxiv2016} derived
from the squared norm of the probability vector $| P(t) |^2$. The
formalism proposed therein allow us to further explore the operatorial
content of the corresponding Markov process. More specifically, we
demonstrate the time evolution is also achieved in a framework that
only requires eigenvalues derived from Hermitian operators. This leads
to a constrained multivariate equation, which can be solved by standard
optimization techniques. We explore the fact the time evolution heavily
relies on the eigendecomposition and formulate perturbative corrections
to epidemic models. Additionally, we also discuss methods usually employed
in quantum many-body problems and Statistical Physics, such as the the
Bethe-Peierls approximation \cite{betheProcRoySoc1935} and the
Holstein-Primakoff transformations
\cite{primakoffPhysRev1940,emaryPhysRevE2003,emaryPhysRevLett2003}. 


 \section{SIS model}
 \label{sec:trans_matrix}

\begin{figure}
\centering
\includegraphics[width=0.5\textwidth]{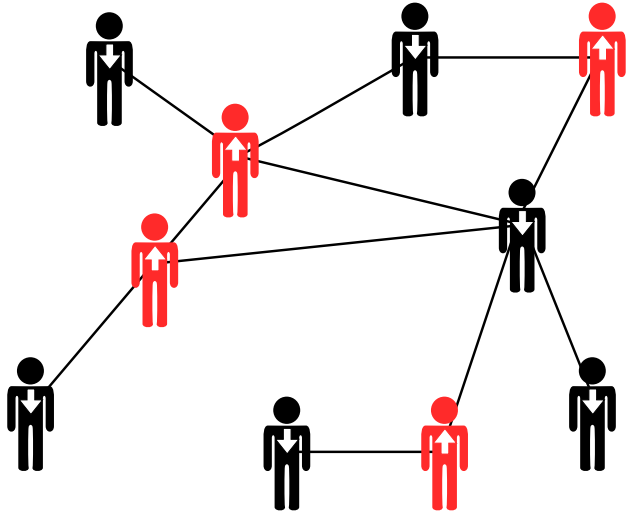}
\caption{\label{fig1} Networks are used to reproduce the heterogeneous
  interactions in a population formed by susceptible ($\downarrow$, black)
  and infected agents ($\uparrow$, red). The topology of the network figures
  among the relevant factors that impact the disease spreading.}
\end{figure}

One of the key aspect of agent-based models is the use of networks to
describe heterogeneity among the distinct agents
\cite{keelingPNAS2002,keelingJRSoc2005,bansalJRSoc2007,satorrasRevModPhys2015},
as Fig.~\ref{fig1} depicts. By taking into account the individuality of
each agent, the web of connections between them creates disease
spreading patterns, due to the development of characteristic pathogen
mobility within the underlying population. Epidemic time duration is
shorter for populations consisting of loose connected agents, whereas
the potential of disease dissemination is expected to be stronger for
hub-like agents.  

Networks are characterized by their topological quantities such as
degree, connectivity, centrality, \emph{etc.}, and we label the
collection of these descriptive quantities as $\omega$. As long as they
share the same set of topological characteristic values $\omega$,
several distinct objects may in fact  represent the same network
$\mathcal{N}(\omega)$.  Graphs are natural candidates to represent
networks~\cite{albertRevModPhys2002} since they are mathematical
constructions formed by interconnected vertices $V_k$ ($k= 0, \ldots,
N-1$).  For each graph, the adjacency matrix $A$ ($N\times N$) describes
all present connections among vertices; the matrix elements are $A_{i
  j}=1$, if vertices $i$ and $j$ are connected or vanish otherwise.  In
this context, an ensemble of graphs is a convenient way to represent a
network, \emph{i.e.}, one graph $G(\omega)$ is a single
realization of the network $N(\omega)$, whose set of topological
quantities are $\omega$.

The correspondence between networks and agent-based models requires that
each vertex contains exactly one agent, which state belongs a discrete
domain.  More precisely, $V_k$ holds the discrete state $k$-th agent and
$A$ reproduce the connection among agents.  Once agents and their
interconnections are properly written, we address the $N$-agent
configuration health state.  Let the current health status of the $k$-th
agent be $\sigma_k$, which for may acquire two values, namely,
$\sigma_k = \downarrow$ (susceptible) or $\sigma_k = \uparrow$
(infected).  The vector 
\begin{equation}
\lvert C_{\mu} \rangle \equiv \lvert
\sigma_1 \sigma_2 \cdots \sigma_N \rangle,
\end{equation}
with $\mu = 0, 1, \ldots, 2^N - 1$, describes the health status
configuration of $N$ agent. Since there are two health states
available per agent, the total number of distinct configurations is
$2^N$. We enumerate each configuration $C_{\mu}$ using binary
arithmetic:  
\begin{equation}
\mu = \delta_{\sigma_0,\uparrow} 2^0 +  \delta_{\sigma_1,\uparrow} 2^1 +
\cdots + \delta_{\sigma_{N-1},\uparrow} 
2^{N - 1}.  
\end{equation}
For instance, the configuration containing only healthy individuals
is $\lvert C_0\rangle = \lvert \downarrow\downarrow\downarrow\cdots
\downarrow\rangle$, whereas the configuration where only the $k=1$
agent is infected is $\lvert C_2\rangle =\lvert
\downarrow\uparrow\downarrow\cdots \downarrow\rangle$. Henceforth, we 
set the following notation: Latin integer indices run over agents
$[0,N-1]$, while Greek integer indices enumerate configurations $[0,2^N-1]$.

The set formed by configurations $\{C_{\mu}\}$ spans the finite vector
space $\mathbbm{H}$. Within $\mathbbm{H}$, one may define the relevant
operators for epidemic spreading. The operator $\hat{\sigma}^z_k$
shows whether the $k$-th agent is infected ($\uparrow$) or not
($\downarrow$), namely,
\begin{equation}
  \hat{\sigma}^z_k | \sigma_1 \sigma_2 \cdots \sigma_N \rangle = (
  \delta_{\sigma_k \uparrow} - \delta_{\sigma_k \downarrow}) | \sigma_1
  \sigma_2 \cdots \sigma_N \rangle.
\end{equation}
The number of infected agents at vertex $k$ is obtained \emph{via} the
operator
\begin{equation}
{\hat{n}_k = \frac{1}{2} \; (\hat{\sigma}^z_k + 1)} \; , 
\end{equation}
while the total number of infected agents in the population is
\begin{equation}
\hat{n}=\frac{1}{2}\sum_{k=0}^{N-1}\hat{\sigma}^z_k+\frac{N}{2}.
\end{equation}
Agent health status are flipped by the action of $\hat{\sigma}_k^+$
and $\hat{\sigma}_k^-$: 
\begin{align}
  \hat{\sigma}_k^+ \lvert\sigma_1\cdots \downarrow_k\cdots\sigma_N
  \rangle =& \lvert \sigma_1\cdots \uparrow_k \cdots\sigma_N\rangle, \\
  \hat{\sigma}_k^- \lvert \sigma_1\cdots \uparrow_k
  \cdots\sigma_N\rangle =& \lvert \sigma_1\cdots \downarrow_k
                           \cdots\sigma_N\rangle, 
\end{align}
null otherwise. They are combined to form another $\hat{\sigma}_k^x
=\hat{\sigma}_k^++\hat{\sigma}_k^-$. The localized operators
$\hat{\sigma}_k^{\pm}$ and $\hat{\sigma}_k^z$ satisfy well-known
algebraic relations. For each $k$, $\hat{\sigma}_k^{\pm, z}$ form
local $\textrm{su}(2)$ algebra, with the following structure constants: 
$  [ \hat{\sigma}_k^z, \hat{\sigma}_k^{\pm}]  =  \pm 2
  \hat{\sigma}_k^{\pm}$ and $ [\hat{\sigma}_k^+, \hat{\sigma}_k^-]  =
  \sigma_k^z$. However, they also exhibit local fermionic
  anticommutation rules, $\{ \hat{\sigma}_k^+, \hat{\sigma}_k^- \}  =
  1$, and non-local bosonic relations, 
$[\hat{\sigma}_k^{r}, \hat{\sigma}_{k'}^{s}] = 0$, for $k\neq k'$ 
and $r,s = \pm, z$. The dual fermionic-bosonic behavior is a
familiar occurrence in quantum spinchains
\cite{alcarazAnnPhys1988,nakamuraPhysicaA2016}. Usually, it is advisable
to select either the fermionic behavior \emph{via} the Jordan-Wigner
transformation \cite{ortizPhysRevLett2001} or, alternatively, the
bosonic behavior \emph{via} the Holstein-Primakoff transformation
\cite{primakoffPhysRev1940}. We postpone the behavior-selection as our
intention in this section concerns general aspects.


For any Markov process, the probability vector represents the system
and is written as
\begin{equation}
\lvert P(t)\rangle =\sum_{\mu} P_{\mu}(t)\lvert C_{\mu}\rangle \; ,
\label{eq:markov_prob}
\end{equation}
where $P_{\mu}(t)$ is the probability to find $N$ agents in the
configuration $\lvert C_{\mu}\rangle$, at time $t$, subjected to
probability conservation constraint,
\begin{equation}
\sum_{\mu}P_{\mu}(t)=1.
\end{equation}
Another integral part of the Markov process is the transition matrix
$\hat{T}$, which weight transitions among configurations in a fixed time
interval $\delta t$, creating the temporal succession: 
\begin{equation}
\lvert P(t+\delta t)\rangle = \hat{T}\lvert P(t)\rangle.
\label{eq_p}
\end{equation} 
The details concerning disease transmission or recovery in epidemic
model are included in $\hat{T}$ by considering operators that act over
the $N$-agent configurations $\lvert C_{\mu}\rangle$.  In the SIS model,
an infected agent at vertex $k$ is subjected to three distinct outcomes
after the action of $\hat{T}$: transmit the disease to a susceptible
connected agent; recover to the susceptible state; or remain unchanged.

The recovery event for the $k$-th agent is executed by the operator
$\hat{\sigma}^{-}_k\hat{n}_k$. Brief inspection shows the action is
quite simple: if the $k$-th agent is currently infected, the health
status flips to susceptible. On contrary, if the $k$-th agent is
susceptible, it returns the null vector.  Although the recovery event
does not involve the underlying network, the disease transmission
event does.  As a result, the corresponding operator $A_{k
  m}\hat{\sigma}^{+}_m \hat{n}_k$ transmits the disease from the $k$-th
agent to $m$-th agent.  Similarly to the recovery process, the operator
$\hat{n}_k$ only checks whether the $k$-th agent is infected.  The
difference appears due to the adjacency matrix $A_{m k}$ and the
infection operator $\hat{\sigma}^+_m$.  Note that if the $m$-th agent is
already infected, the operatorial action returns the null vector as
well.  Finally, the event to remain unchanged requires diagonal
operators and accounts for all the other possible non-diagonal events.
The operator which provides the number of available outcomes of disease
transmission for $k$-th agent is $\sum_{j}A_{j
  k}(1-\hat{n}_j)\hat{n}_k$; whereas the operator that accounts for the
chance to not recover is simply $\hat{n}_k$.

Under the Poissonian assumption, one only considers a single recovery
 or a single infection event, per time interval $\delta t$. Under
this circumstances, the  transition matrix reads 
\begin{equation}
  \hat{T}=\mathbbm{1}- \frac{\alpha}{N}\sum_{k j} \left[   A_{j k} ( 1
    -\hat{n}_j - \hat{\sigma}^{+}_j) + \Gamma \delta_{k
      j}(1-\hat{\sigma}^{-}_j) \right] \hat{n}_k ,
\label{eqtsis}
\end{equation}
with $\Gamma=\beta N/ \alpha$. 
The diagonal components are the probabilities for the configuration to
remain unchanged after one time step. Disease spreading explicitly
carries the network topology due to the contribution of $A$. More
importantly, the transition matrix $\hat{T}$ contains non-Hermitian
operators and, hence, its left and right eigenvectors, $\langle
\chi_{\mu} \rvert$ and $\lvert \phi_{\mu}\rangle$, respectively, are not
related by Hermitian transposition.

We emphasize that the construction of Eq.~(\ref{eqtsis}) considers only
a single graph.  In general, agent-based models are built under the
hypothesis  of $N\gg 1$.  The reasoning behind this choice lies in the
network averaging process.  If the graph is large enough $N\gg 1$, one
expects to recover the network topological quantities $\omega$ within a
single realization.  This statement is the equivalent to the ergodic
hypothesis, where the ensemble average over $M$ graphs is replaced by
the average within a single graph (self-averaging/annealing case).  This
expectation is reasonable but cannot hold for moderate $N$.  In what
follows, we explicitly consider the network ensemble containing $M>1$
graphs (quenched).  For that purpose, let
$G_{i}\in\{G_0,G_1\ldots,G_{M-1}\}$ be the $i$-th graph in the network
ensemble.  Furthermore, for each graph $G_{i}$ there is a corresponding
adjacency matrix $A^i$.  For fixed initial condition, Eq.~(\ref{eq_p})
tells us the time evolution is a linear transformation so the average
over the network ensemble is estimated by $ \hat{\bar{T}} = M^{-1}\sum_i 
\hat{T}^{(i)}$, where $\hat{T}^{(i)}$ is the transition matrix using
the graph $G_i$ ($i=0,1,\dots,M-1$). For the SIS model, the averaging
process is tracked down to 
\begin{equation}
\bar{A}_{j k} = \frac{1}{M}\sum_{l=0}^{M-1} A_{j k}^{(l)},
\label{eq_mean_a}
\end{equation}
with $A^{(l)}$ corresponding to the adjacency matrix of $G_l$ and the
bar symbol represents the ensemble average. Notice that $\bar{A}_{j k}$
is not restricted to the integers 0 or 1 any longer. In practice,
Eq.~(\ref{eq_mean_a}) claims the estimator $\bar{A}$ is a real $N \times
N$ matrix and we can safely drop to bar symbol. However, sudden changes
in the network topological properties $\omega$ must be investigated
using Eq.~(\ref{eq_mean_a}), from which one derives perturbative
operators.

Since $\hat{T}$ in Eq.~(\ref{eqtsis}) is time independent, the Taylor
expansion of Eq.~(\ref{eq_p}) leads to the following system of
differential equations:
\begin{equation}
\frac{d P_{\mu}}{d t} = -\sum_{\nu} H_{\mu \nu}P_{\nu}(t),
\label{eqdinamica}
\end{equation} 
where $H_{\mu\nu}$ are the matrix elements of the time generator
\begin{equation}
{\hat{H}\equiv \frac{\mathbbm{1}-\hat{T}}{\delta t}} \; . 
\label{eqdefh}
\end{equation}
The operator $\hat{H}$
governs the dynamics of disease spreading and whose normal modes are
labeled after the eigenvalues $\{ \lambda_{\mu} \}$. The formal solution
to Eq.~(\ref{eqdinamica})  is
\begin{equation}
\lvert P(t)\rangle = \mathe^{-\hat{H} t}\lvert P(0)\rangle.
\label{eqtimeevolP}
\end{equation}
Clearly, the eigenvalues must satisfy $\lambda_{\mu}\geq 0$, for any
$\mu$, vanishing only for stationary states \cite{reichl1998}. In
addition, the corresponding spectral density function $\rho(\lambda)$
depends on the couplings present in  Eq.~(\ref{eqtsis}), namely, the
disease transmission and recovery probabilities as well as the network
average adjacency matrix $A$.

\section{Theoretical approach}

One of the main goals in epidemic models is the development of methods to
predict the way epidemics change when parameters are subjected to small
variations. If such predictions are robust, preemptive actions to lessen
the epidemic are also expected to achieve better results. In physical
theories, small changes in couplings or substrate are often investigated
under perturbative schemes based on simpler models, which often have
known orthonormal modes.  In epidemic models, however, one must work
with asymmetric operators $\hat{H}$ and their left and right
eigenvectors in Eq.~(\ref{eqtimeevolP}). 
Despite the complexities related to the operatorial content
of $\hat{H}$, the stochastic nature of the problem ensures the
conservation of total probability $\sum_{\mu}P_{\mu}(t)=1$ for any
$t$. Another relevant descriptive variable derived from $\lvert
P(t)\rangle$ is the squared norm,
\begin{equation}
\lvert P(t)\rvert^2 = \langle P(t)\vert P(t)
\rangle=\sum_{\mu=0}^{2^N-1}P_{\mu}(t)^2.
\label{eqnorma2}
\end{equation}
As noted in Ref.~\cite{nakamuraArxiv2016}, probability conservation does
not enforce conservation of $\lvert P(t)\rvert^2$ along time, occurring
only after the system reaches complete stationarity. Therefore, $\lvert
P(t)\rvert^2$ may be used to assess general properties of the stochastic
model during both transient and stationary phases.

\begin{figure}[hb]
\centering
\includegraphics[width=0.5\columnwidth]{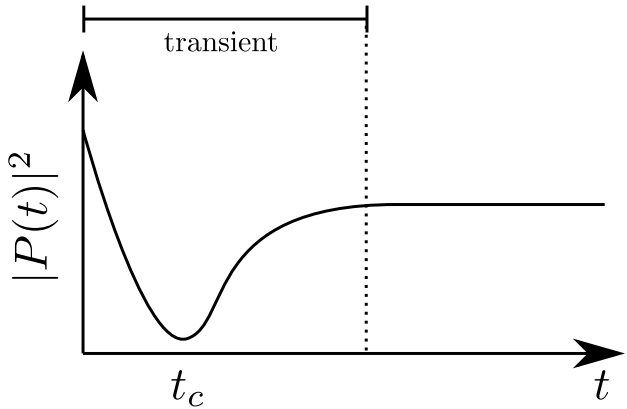}
\caption{\label{fig2} Time evolution of $\lvert P(t)\rvert^2$ for
  arbitrary Markov process. The system starts at $t=t_0$ with initial
  condition $P_{\mu}(t_0)=\delta_{\mu\xi}$. During the transient phase, 
  the number of available configurations increases, lowering the value
  of $\lvert  P (t)\rvert^2$. At $t=t_c$,  $\lvert  P (t_c)\rvert^2$
  develops a minimum.  As the system continues to evolve,
  the squared norm continuously increases until stationarity is
  achieved. In this particular illustration, the number of stationary
  states is larger than unity, producing
  $\lim_{t\rightarrow\infty}\lvert  P (t)\rvert^2 < 1$.} 
\end{figure}

Since the squared norm can only assume values in the interval $[0,1]$,
one may consider a single differential equation to investigate 
the time evolution of $\lvert P(t)\rvert^2$. Taking the time derivative
of  Eq.~(\ref{eqnorma2}) and using Eq.~(\ref{eqdinamica}) results in 
\begin{equation}
-\frac{d}{dt} \vert P(t)\vert^2 =
2\langle P(t)\rvert \hat{\mathcal{H}} \lvert P(t)\rangle.
\label{eqsqnorm}
\end{equation}
 Unlike $\hat{H}$, the symmetrized time generator
 \begin{equation}
{\hat{\mathcal{H}}\equiv \frac{\hat{H}^T+\hat{H}}{2}}
\end{equation}
is Hermitian with orthonormal basis $\{\lvert \psi_{\mu}\rangle\}$ and
corresponding eigenvalues $\{\Lambda_{\mu}\}$, for $\mu=0,\ldots,2^N-1$. 
The eigenvalues $\Lambda_{\mu}$ differ from
their counterparts $\lambda_{\mu}$, since $\Lambda_{\mu}$ are not
positive semi-definite, while the complex coefficients 
\begin{equation}
 \pi_{\mu}(t) \equiv  \langle \psi_{\mu}\vert P(t)\rangle 
\end{equation}
are not probabilities, even though they are used to evaluate the
configurational probabilities
\begin{equation}
P_{\mu}(t)=\sum_{\nu}\langle C_{\mu}\vert \psi_{\nu}\rangle\, \pi_{\nu}(t) .
\end{equation}
Figure~\ref{fig2} illustrates the time evolution of $\lvert P(t)\rvert^2$ for
arbitrary Markov process.

The spectral decomposition of $\lvert P(t)\rangle
=\sum_{\mu}\pi_{\mu}(t)\lvert \psi_{\mu}\rangle$ in Eq.~(\ref{eqsqnorm}) produces:
\begin{equation}
\sum_{\mu}\left(\frac{1}{2}\frac{d}{d t}+ \Lambda_{\mu} \right)\lvert\pi_{\mu}(t)\rvert^2=0,
\label{eqeig1}
\end{equation}
subjected to the constraint $\sum_{\mu\nu}\langle C_{\mu} \vert
\psi_{\nu} \rangle\pi_{\nu}(t)=1$. Notice that Eq.~(\ref{eqeig1}) is
valid for any time instant $t$. As such, one may also integrates the
expression in Eq.~(\ref{eqeig1}) taking into account the probability
constraint with help of one Lagrange multiplier $h$,
\begin{align}
\mathcal{S}_0& \equiv\sum_{\mu}\int_{t_0}^{t_f} dt
               \left[\frac{1}{2}\frac{d}{d t} \lvert\pi_{\mu}\rvert^2+ 
               \Lambda_{\mu} \lvert\pi_{\mu}\rvert^2 \right],
               \label{eqeig21}
  \\
  \mathcal{S}&\equiv S_0 + h\sum_{\mu\nu}
               \int_{t_0}^{t_f}dt \left[ \frac{\pi_{\nu}}{2}\langle C_{\mu}|
               \psi_{\nu}\rangle  + \frac{\pi_{\nu}^*}{2}\langle \psi_{\nu}|
               C_{\mu}\rangle  -1\right],
               \label{eqeig22}
\end{align}
where $t_0$ and $t_f$ are the fixed initial and final time instants,
respectively. Equations~(\ref{eqeig21}) and (\ref{eqeig22}) share
striking similarity with the classical action of Mechanics
\cite{goldsteinClassicalMechanics}. 
Since we are only interested in stationarity and extrema states
so we can neglect Eqs.~(\ref{eqeig21}-\ref{eqeig22}) for now. In fact,
the condition of vanishing derivative in Eq.~(\ref{eqeig1})
results in the following algebraic equation:
\begin{equation}
\sum_{\mu}|\tilde{\pi}_{\mu}|^2\Lambda_{\mu}=0,
\label{stat5}
\end{equation} 
where $\tilde{\pi}_{\mu}$ either are the coefficients corresponding to
stationary states or local extrema.  

Formally, Eq.~(\ref{stat5}) may be solved using optimization
algorithms for constrained quadratic problems. Of course, $2^N$
optimization problems are still formidable numerical problems and also
involves the numerical approach taken for each optimization algorithm as
well. Failure to converge to correct solutions or only walk in a
particular neighborhood in the solution space are among common sources
of problems. Furthermore, the derivation of Eqs.~(\ref{eqeig1}) and
(\ref{stat5}) assumes the eigenvalues $\{\Lambda_{\mu}\}$ and the
corresponding eigenvectors $\{\psi_{\mu}\}$ are known, which again may
be a complex $2^N$ diagonalization problem depending on the algebraic
form of $\hat{\mathcal{H}}$.

However, the aforementioned hardships are the crucial aspects one must
consider to decide whether to use Eqs.~(\ref{eqeig1}) and (\ref{stat5})
or Eq.~(\ref{eqdinamica}).  The answer is very simple: $\lvert
P(t)\rvert ^2$ is only useful if symmetries are present in
$\hat{\mathcal{H}}$ but not in $\hat{H}$ \cite{nakamuraArxiv2016b}.
Additional symmetries 
simplify the diagonalization problem and also introduce explicit bounds  
in root-finding procedures. It is easy to find examples where such
symmetry increase occurs. For instance, consider $\hat{H}=
\hat{\sigma}^+_1 +\hat{\sigma}^+_2$ so that the corresponding Hermitian
generator is $\hat{\mathcal{H}}= \sigma^{x}_1 +\sigma^{x}_2$, whose
eigenvectors are grouped according to the quantum number
$m_x=-1,0,1$. Therefore, if additional symmetries are available for
$\hat{\mathcal{H}}$, traditional optimization algorithms become valuable
resources to solve Eq.~(\ref{stat5}) and, thus, the stationary
states of Markov processes and their corresponding occurrence
probabilities.

As a practical example to verify the results in Eq.~(\ref{stat5}), 
consider the SIS model with $N=3$ agents and 
$\alpha=N/10$ and $\Gamma=0$, in the fully connected network $A_{i
  j}=(1-\delta_{ij})$. This set of parameters and network topology
reproduce the SI model. From Eqs.~(\ref{eqtsis}) and (\ref{eqdefh}), the
Hermitian time generator for the  SIS model is  
\begin{align}
\hat{\mathcal{H}}_{\textrm{SIS}}=&\hat{\mathcal{H}}_0 + \hat{\mathcal{H}}_1 \; ,\\
  \hat{\mathcal{H}}_0=&+\frac{\alpha}{N}\sum_{k j} \left[   A_{j k} ( 1
    -\hat{n}_j) + \Gamma \delta_{k       j}\right] \hat{n}_k ,\label{eqsish0}\\
  \hat{\mathcal{H}}_1=&-\frac{\alpha}{2N}\sum_{k j} \left[   A_{j k} ({
                       \hat{\sigma}_j^+ \hat{n}_k + \hat{n}_k
                       \hat{\sigma}^-_j}) +{\Gamma}\delta_{k 
                       j} \hat{\sigma}_k^x
                       \right]. \label{eqsish1}
\end{align}
 In this case, there are four eigenvalues of
$\hat{\mathcal{H}}_{\textrm{SIS}}$ relevant to the description of
stationary states, namely, $\Lambda_0=0$, $\Lambda_3=0.1571993$,
$\Lambda_6=0.3514137$ and $\Lambda_7=-0.1086130$. The trivial stationary
state (none infected) is obtained setting $\tilde{\pi}_{\mu}=\delta_{\mu
  0}$. The stationary state  where all agents are infected is obtained
using $\tilde{\pi}_{3}=0.3977703$, $\tilde{\pi}_{6}=-0.3803660$ and
$\tilde{\pi}_{7}=0.8349255$.

\section{Perturbative methods and approximations}

The ability to predict causal effects in the disease spreading is surely
desirable for any epidemic model, as it allows decision-makers to select
adequate strategies to mitigate new incidence cases and funding
priorities. Among them, effects caused by small perturbations in the
underlying contact network are specially important for agent-based
models. As Ref.~\cite{satorrasRevModPhys2015} states, heterogeneous
population hinders analytical insights about perturbative effects.
Nonetheless, Eq.~(\ref{stat5}) provides an alternative way to introduce
perturbative methods and approximations to epidemic models, since it
relies on the Hermitian generator $\hat{\mathcal{H}}$. This is relevant
because the standard time independent perturbation theory may be used to
evaluate corrections to the quantities relevant to Eq.~(\ref{stat5}),
namely, the eigenvalues $\Lambda_{\mu}$ and the coefficients
$\tilde{\pi}_{\mu}$.

For the sake of clarity, we consider a finite number of topological values
$\omega_r$ ($r=1,\ldots,R$) to characterize the network.
In this context, a simple perturbative scheme is attained by
considering the change $\omega_r \rightarrow \omega_r+\delta\omega_r$,
with $|\delta\omega_r| \le \delta\omega$ for any 
$r=1,2,\ldots,R$ and fixed $\delta\omega\ll 1$. The quantities $\omega_r$ are
expected to be complicated functions so that the  variations
$\delta\omega_r$ are not completely independent. Nonetheless, they are
still calculated from estimators derived from the average adjacency
matrix $\bar{A}$. Therefore, one expects a corresponding perturbative
matrix $\delta\omega B$ to be added to the average adjacency matrix:
\begin{equation}
\bar{A}_{ij}\rightarrow \bar{A}_{ij}+\delta\omega B_{i j}.
\label{eqpert1}
\end{equation}
The details of the matrix representation $B$ are specific for
each perturbation set $\{\omega_r\}$ adopted, but the relevant
information lies in the coupling $\delta\omega$, as it
provides a natural perturbative parameter.
Now, it is a simple task to identify the perturbation $\hat{V}$ in the
time generator, 
\begin{align}
  \hat{\mathcal{H}}&=\hat{\mathcal{H}}_0+\hat{\mathcal{H}}_1+\delta\omega\,
                     \hat{V},\\
  \hat{V}&=\frac{\alpha}{2N}\sum_{k j}B_{j k}\left[ 2(1-\hat{n}_j)
           \hat{n}_k
           -\hat{\sigma}_j^+\hat{n}_k-\hat{n}_k\hat{\sigma}_j^-
           \right].
           \label{eqv}
\end{align}
A few selected cases merit further attention. First, the special case
$B=z A$, with $\delta\omega z\in [-1,1]$, recovers the effective
coupling formulation $\alpha\rightarrow \alpha(1+z\delta\omega)$ in
random networks. Another interesting case occurs if $A$ and $B$ are
periodic regular networks with distinct periods, $t_A\neq t_B$,
respectively. Depending on the initial condition and the ratio
$t_B/t_A$, the perturbation $\hat{V}$  may either connect all available
states, or lock the time evolution in a periodic cycle.

Hermiticity is sufficient to warrant Rayleigh-Schr\"odinger perturbation
theory and produces first order corrections to eigenvalues and
eigenvectors, respectively, 
\begin{align}
\Lambda_{\mu}^{(1)}&=\langle \psi_{\mu} \vert\hat{V}
\vert \psi_{\mu}\rangle,
\label{eqeigcorr1}\\
\tilde{\pi}_{\mu}^{(1)}&=\sideset{}{'}\sum_{\nu}\frac{\lvert \langle
\psi_{\mu}\vert\hat{V} \vert \psi_{\nu}\rangle
\rvert^2 }{\Lambda_{\nu}-\Lambda_{\mu}}.
\label{eqeivcorr1}
\end{align}
The prime indicates the sum excludes degenerate states with eigenvalue
$\Lambda_{\mu}$. Substituting these results into Eq.~(\ref{stat5}) and
discarding second order corrections, one arrives at
\begin{equation} 
  \sum_{\mu}\left[ 2 \Lambda_{\mu} \textrm{Re}(\tilde{\pi}_{\mu}\tilde{\pi}_{\mu}^{(1)})+
    \lvert\tilde{\pi}_{\mu}\lvert^2\Lambda_{\mu}^{(1)}
\right]=0.
\label{stat6}
\end{equation}
While the perturbative corrections are given by Eqs.~(\ref{eqeigcorr1})
and (\ref{eqeivcorr1}), the relation in Eq.~(\ref{stat6}) shows they
might not be independent.  Similarly, the perturbative corrections for
configurational probabilities  $P_{\mu}(t)$ read 
\begin{equation}
P_{\mu}^{(1)}(t)=\sum_{\nu}\pi^{(1)}_{\nu}(t)\langle C_{\mu}\vert
\psi_{\nu}\rangle \; . 
\end{equation}

Alternatives to perturbation theory are readily available as well, since
the only requirement for Eq.~(\ref{stat5}) are eigenvalues and
eigenvectors of $\hat{\mathcal{H}}$. This means analytical and numerical
techniques, usually available only for quantum many-body theories, are
now available to epidemic models such as the Bethe-Peierls  meanfield
approximation (BPA) \cite{betheProcRoySoc1935} and the
bosonification~\cite{primakoffPhysRev1940}.  

In the BPA, the operator $\hat{n}_k$ is replaced by global average
$\bar{n}$, producing the effective time generator  
\begin{equation}
\frac{\hat{\mathcal{H}}'}{\alpha/N}= \frac{\Gamma
  N}{2}+\frac{\bar{n}}{2}\sum_j\kappa_j+\frac{1}{2}\sum_j\left[\Omega_j({\cos\theta_j
    \hat{\sigma}^z_j-\sin\theta_j \hat{\sigma}^x_j})\right],
\label{effh}
\end{equation}
where $\kappa_j=\sum_k \bar{A}_{k j}$ is the degree of $j$-th vertex,
$\Omega_j=\sqrt{2(\Gamma^2+\bar{n}^2 \kappa_j^2)}$, $\cos\theta_j =
(\Gamma-\bar{n}\kappa_j)/\Omega_j$ and $\sin\theta_j =
(\Gamma+\bar{n}\kappa_j)/\Omega_j$.  
The effective generator $\hat{\mathcal{H}}^{\prime}$ in
Eq.~(\ref{effh}) is diagonalized by rotations around the $y$-axis:
\begin{align}
\Lambda_{\mu}' &=  \frac{\Gamma N}{2}+\frac{\bar{n}}{2}\sum_j\kappa_j+
\frac{1}{2}\sum_{j} \Omega_j (-1)^{m_j},\\
\mu &= m_02^0+m_12^1+\cdots+m_{N-1}2^{N-1}.
\end{align}
Due to the main requirement $\hat{n}\rightarrow \bar{n}$, the BPA rules out its
applicability during transient. For stationary states, however, the BPA
provides a convenient coarse particle picture accompanied by all
eigenvalues and corresponding eigenvectors.

Symmetries are crucial ingredients to reduce the complexity associated
with the spectral decomposition in quantum many-body problems.
In what follows, we investigate finite networks with permutation
invariance to better understand the role played by finite symmetries. This is
possible due to Cayley's theorem \cite{tinkham}.  In the SIS model, the
fully connected network $A_{i    j}=(1-\delta_{ij})$ exhibits the
desired symmetry. This simple case is used  as training grounds for
non-trivial networks. The first step is to unravel the role played by
quantum angular operators   
\begin{align}
\hat{J}^{\pm}&=\sum_{k}\hat{\sigma}^{\pm}_k,\label{eqjpm}\\
\hat{J}^z&=\sum_k\hat{\sigma}^z_k-\frac{N}{2}.
\label{eqjz}
\end{align}
The algebraic relations are $[\hat{J}^+,\hat{J}^-]=2\hat{J}^z$ and
$[\hat{J}^z,\hat{J}^{\pm}]=\pm\hat{J}^{\pm}$ so that $\hat{J}^{\pm,z}$
form a compact Lie algebra with  Casimir operator
 $\hat{J}^2=(\hat{J}^z)^2+(1/2)\{\hat{J}^+,\hat{J}^-\}$.

The SIS symmetric time generator $\hat{\mathcal{H}}_{\textrm{SIS}}$ is
obtained from Eqs.~(\ref{eqsish0}) and (\ref{eqsish1}) and expressed
using Eqs.~(\ref{eqjpm}) and (\ref{eqjz}):
\begin{equation}
  \hat{\mathcal{H}}_{\textrm{SIS}}=
  \frac{\alpha}{N}\left( N-\hat{n}+\Gamma \right) \hat{n}
  -\frac{\alpha}{2N}\left[ \hat{J}^{+}\hat{n}+\hat{n}\hat{J}^- +\Gamma
    \left({\hat{J}^++\hat{J}^-}\right)\right]. 
\label{eqhsis3}
\end{equation} 
It
 should be noted the appearance of global angular operators $J^{\,\pm}$
 is a direct consequence of the network choice adopted here, as it
 captures important global properties, including rotations.  For
 non-trivial network  topologies, one must consider localized angular
 momentum operators  $\hat{J}^{\,\pm}_k$ as usual in many-body problems.    
A brief inspection of Eq.~(\ref{eqhsis3}) shows 
\begin{equation}
[\hat{\mathcal{H}}_{\textrm{SIS}},\hat{J}^2]=0,
\end{equation}
meaning the eigenvalues $j(j+1)$ are conserved quantities and the number of
infected agents may only assume the following values $n=0,1,\ldots,2j$
for fixed value $j$. 
Under these circumstances, one introduces the Holstein-Primakoff
transformations, which exchange the set of angular operators for the
harmonic oscillator destruction and creation operators, $\hat{a}$ and
$\hat{a}^{\dagger}$, respectively, with $[\hat{a},\hat{a}^{\dag}]=\mathbbm{1}$. 
The transformations for the $j=N/2$ sector are   
\begin{align}
\hat{J}^+=&\;\;\sqrt{N+1-\hat{n}} \;\hat{a}^{\dag},\label{eqhp1}\\
\hat{J}^-=&\hat{a} \sqrt{N+1-\hat{n}}  \;\;\;\;.\label{eqhp2}
\end{align}
Usually, the Holstein-Primakoff transformations are most useful when the
average occupation number satisfies $\langle n\rangle/N \ll 1$ for $N\gg 1$. 
Under this condition one may expand the square root and keep the linear order in $\hat{n}/N$. 
Consequently, the analysis of epidemics suggests the employment of coherent states,
\begin{equation}
  \lvert \lambda\rangle =
  \mathe^{-\lambda^2/2}\sum_{m=0}^{\infty}\frac{\lambda^m}{\sqrt{m!}}\lvert m\rangle,
\end{equation}
since they satisfy the eigenvalue equation  $\hat{a}\lvert \lambda\rangle = \lambda
\lvert \lambda\rangle$ with $\langle \lambda\vert
\hat{n}\vert\lambda\rangle=\lambda^2$. Here, we only consider 
$\lambda\in\mathbbm{R}$. Another remarkable property of coherent states
is that several observables are derived from the Poisson distribution.
 Under this scheme,
\begin{align}
\frac{\langle \lambda \vert \hat{\mathcal{H}}_{\text{SIS}}\vert \lambda\rangle}{\alpha/N}
  &= \lambda^2 \left[X_N(\lambda^2) +(N-1) Y_{N-1}(\lambda^2)\right]
  -\lambda Z_N,\label{eqhsis_coherent}\\
  X_N(\lambda^2)&=\frac{\Gamma_{N-1}(\lambda^2)}{\Gamma_{N-1}(0)}(N-1-\lambda^2)+
                  \frac{\Gamma_N(\lambda^2)}{\Gamma_{N}(0)} \Gamma,\\
  Y_N(\lambda^2)&=\frac{\lambda^{2N} \mathe^{-\lambda^2}}{N!},\\
  Z_N(\lambda^2)&=\sum_{m=0}^N{\left[\frac{\mathe^{-\lambda^2}\lambda^{2m}}{m!}\sqrt{N-m}\left(m+\Gamma\right)\right]}
\end{align}
and $\Gamma_{m}(\lambda^2)$ is the incomplete Gamma function for
integer $m$. 
As a closing remark, we emphasize the rationale behind this approach: since coherent states are eigenvectors of $\hat{a}$, they remain unchanged under successive actions of $\hat{\mathcal{H}}_{\text{SIS}}$, making them suitable candidates to characterize epidemics as $t\rightarrow\infty$.


\section{Conclusion}

Fluctuations are integral part in stochastic processes. In compartmental
approaches to epidemics, their role are underestimated when the
population of infected agents is scarce and heterogeneous. However,
disease spreading models describing Markov processes are limited to
small population size due to asymmetric time generators, which hinder
the development of novel analytical insights. This issue is addressed by
employing  $\lvert P(t) \rvert^2$, which provides a single multivariate
equation, requiring only eigenvalues and eigenvectors of symmetric time
generators. One way to exploit this fact is to divert efforts in solving
the symmetric spectral decomposition. 
Our finding shows the development of a perturbation theory in epidemic models, with emphasis in the aspects produced by topological perturbations, as show in Eq.~(\ref{eqv}).
In addition, the Bethe-Peierls approximation provides the complete
eigenspectrum and eigenvectors. Within this approximation, one
focus in particle-like normal modes. Finally, the Holstein-Primakoff
transformations exploits the network rotation symmetry to uncover 
a framework with quantum oscillators. From this result, one concludes
coherent states are suitable candidates to study large scale
epidemics. Since coherent states can also be used to study losses, we
expect them to provide further hints about the epidemic decay times.


\begin{acknowledgments}
We are grateful for T. J. Arruda and F. Meloni comments during
manuscript preparation.  A.S.M acknowledges grants CNPq 485155/2013 and
CNPq 307948/2014-5.  
\end{acknowledgments}




%

\end{document}